\documentclass[%
 reprint,
superscriptaddress,
 amsmath,
 amssymb,
 aps,
 amsthm
]{revtex4-2}

\usepackage{graphicx}
\usepackage{subcaption}
\usepackage{braket}
\usepackage{tabularx}
\usepackage{booktabs}
\usepackage{xcolor}
\usepackage{amsmath,amssymb}
\usepackage{dcolumn}
\usepackage{bm}
\usepackage{hyperref}
\usepackage{mathtools}
\usepackage{array}
\usepackage{booktabs}
\usepackage{caption}

\begin{document}

\preprint{APS/123 -QED}

\title{Sum-over-poles framework for quantum embedding to \\approach the real-frequency axis}

\author{Alessandro Carbone}
\affiliation{Theory and Simulation of Materials (THEOS), and National Centre for Computational Design and Discovery of Novel Materials (MARVEL), École Polytechnique Fédérale de Lausanne, 1015 Lausanne, Switzerland}
\author{Massimo Capone}
\affiliation{International School for Advanced Studies (SISSA), via Bonomea 265, 34136 Trieste, Italy}
\author{Nicola Marzari}
\affiliation{Theory and Simulation of Materials (THEOS), and National Centre for Computational Design and Discovery of Novel Materials (MARVEL), École Polytechnique Fédérale de Lausanne, 1015 Lausanne, Switzerland}
\affiliation{Paul Scherrer Institut, 5232 Villigen, PSI, Switzerland}
\affiliation{Theory of Condensed Matter, Cavendish Laboratory, University of Cambridge, Cambridge, CB3 0US, United Kingdom}
\author{Tommaso Chiarotti}
\affiliation{Theory and Simulation of Materials (THEOS), and National Centre for Computational Design and Discovery of Novel Materials (MARVEL), École Polytechnique Fédérale de Lausanne, 1015 Lausanne, Switzerland}
\affiliation{Department of Applied Physics and Materials Science, California Institute of Technology,
Pasadena, CA 91125, USA}

\date{\today}

\begin{abstract}
    \noindent
    We introduce a real-frequency framework for the quantum embedding of solids, based on sum-over-poles (SOP) representations. We partition the lattice into periodically repeated interacting fragments and derive an embedded Dyson equation for the fragment, to be solved self-consistently. The resulting scheme generalizes the algorithmic-inversion method of our previous work to interacting Hamiltonians and is closely related to dynamical mean-field theory (DMFT), especially when an extended impurity problem is introduced. 
    By exploiting the compact pole expansion of the embedding potential, we carry out the self-consistent procedure close to the real axis, which avoids ill-posed analytic continuation.
    In addition, the SOP representation allows us to introduce a pole-and-residue dynamics, which we use to reach the self-consistent solution. After deriving the method, we benchmark it against exact-diagonalization implementations of DMFT on the one-dimensional Hubbard ring at half filling. The framework offers a natural route to real-space dynamical quantum embedding, for periodic systems as well as for strongly correlated defects in materials.
\end{abstract}

\maketitle


\section{Introduction}


Understanding the physics of interacting electrons remains one of the central challenges in condensed-matter theory. Many of the most striking phenomena in correlated materials - metal-insulator transitions \cite{imada_1998}, quasiparticle renormalization \cite{schafer_2004}, Kondo screening \cite{sorensen_1996}, unconventional superconductivity \cite{cao_2018}- arise from a subtle interplay of itinerancy and local interactions that cannot be captured within static or mean-field descriptions of the electronic structure. While first-principles frameworks such as density-functional theory \cite{kohn_sham_1965} often provide reliable band structures, their inherent lack of dynamical correlations limits their predictive power when treating strongly correlated materials.

Quantum embedding approaches provide a powerful and systematically improvable route to bridge this gap.
In a nutshell, these approaches are based on the identification of a "fragment" of the whole system, which is ideally solved exactly, while the rest of the system and the interaction between the two parts are treated within some suitable approximation.
For some molecular system of impurities, the natural choice of the fragment is the active part of a larger system, while the rest of the system is weakly correlated.
For a periodic solid, as we show below, an embedding scheme arises by partitioning the system in periodically repeated fragments coupled via a dynamical embedding potential  $v_\text{emb}(\omega)$.

Different embedding schemes differ essentially in the way the properties of the whole system are built from those of the fragment. Influential examples are density-matrix embedding theory (DMET) \cite{Knizia} and its extensions \cite{wouters_2016,sriluckshmy_2021}, quantum defect embedding theory (QDET) \cite{chen_2025}, self-energy embedding theory (SEET) \cite{lan_2017}, and modern formulations of Gutzwiller/slave boson and its ghost extension \cite{lanata_2023,mejutozaera_2024}.
The reference method in the field is dynamical mean-field theory (DMFT) \cite{Georges}, where the Green's function of the lattice is obtained via the knowledge of that of the fragment in the presence of a dynamical Weiss field (or $v_\text{emb}(\omega)$ in this context). The conventional DMFT is based on a single-site fragment, but cluster extensions enlarge the scope to include non-local correlations.


Several recent works have shown that dynamical quantities can be expressed exactly, or to controllable accuracy, as compact rational functions \cite{zhang_minial_2025,roy_1989,huang_2025}. In particular, the algorithmic inversion method\cite{Chiarotti,Chiarotti_2} provides an efficient and stable way to evaluate Green's functions directly on arbitrary complex-frequency paths using a sum-over-poles (SOP) representation. Here, we combine the embedding formulation with a SOP representation of the dynamical embedding potential $v_\text{emb}$, as for a sufficient number of poles, this is an exact representation of time-ordered propagators~\cite{farid_1999}.

\bigskip

In this work we develop an extension of this latter approach to a dynamical quantum embedding formalism based on Green's functions. 
In this context, a self-consistent equation for a generic crystal emerges, when the self-energy is approximated to be local within the fragment.
Then, by representing the dynamical embedding potential as a SOP we map the resulting problem to a fictitious system with an interacting fragment and a non-interacting bath, effectively generalizing the approach to tackle interacting many-body Hamiltonians. 
Specifically, this mapping retains the interacting Hamiltonian on the impurity while introducing an auxiliary Hilbert space based on the SOP representation of a specific frequency-dependent function.
Our construction turns out to be equivalent to the discretized impurity model used in DMFT when an auxiliary Hamiltonian is introduced in order to use exact diagonalization (ED)~\cite{Caffarel_exact_1994,amaricci_2022} or  tensor networks~\cite{nishimoto_2006,wolf_2014,bauernfeind_2017} to compute the Green's function. 
Here, exploiting the SOP formalism, we implement the dynamical quantum embedding framework on real frequencies.
More precisely, to avoid the singularities of time-ordered propagators, all the dynamical quantities are computed on a frequency axis with a small frequency-dependent imaginary part, $z(\omega) = \omega + i \eta \text{erf}(\omega)$.
Compared with conventional implementations based on imaginary axis calculations, this bypasses the numerical issues of analytic-continuation procedures and it is expected to reproduce more faithfully real-frequency spectral features.



\medskip

We test our algorithm on the one-dimensional Hubbard chain by performing both imaginary and real frequency calculations on the same system for different values of the interaction $U$ and different values of $\eta$.
Our results show that compact pole expansions suffice to recover the essential spectral and thermodynamic features of the model, even near challenging correlation regimes. Our approach thus opens the door to real-frequency quantum embedding for realistic electronic-structure Hamiltonians~\cite{bau_first-principles_2026}, multi-orbital problems, and other dynamical embedding strategies beyond standard DMFT.
Importantly, as we also observe below, in this framework, the poles and residues of the embedding potential are determined self-consistently from Eq.~\eqref{eqn:emb_selfcons}.
This paves the way for an algorithm that does not rely on any frequency sampling, but optimizes poles and residues to satisfy this self-consistent condition.

\section{Quantum Embedding on a lattice}
\label{sec:quantum_emb_lat}

We start from the definition of the Hamiltonian of a many-body system on a lattice, its free energy in the grand canonical ensemble and second quantization reads:

\begin{equation}
    \begin{split}
        \hat{K} = \hat{H} - \mu \hat{N} = & \sum_{ij} h_{ij} \hat{c}_i^\dagger \hat{c}_j + \frac{1}{2} \sum_{i,j,k,l} v_{ijkl} \hat{c}_i^\dagger \hat{c}_j^\dagger \hat{c}_l \hat{c}_k  \\
        & - \mu \sum_i \hat{c}_i^\dagger \hat{c}_i\, .
    \end{split}
    \label{eqn:mbHam}
 \end{equation}
 
where $i$ is the lattice index, $\hat{c}^\dagger/\hat{c}$ are the creation/annihilation operators for the single-electron states, $h$ accounts for the kinetic energy term and the electron-nuclei attraction, $v$ is a general electron-electron interaction, and $\mu$ the chemical potential of the system. 
The Dyson equation for the single-particle Green's function of this system is given by:

\begin{equation}
    G^{-1}(\omega) = \omega + \mu - h - \Sigma(\omega) \, .
    \label{eqn:dyson}
\end{equation}

As said in the Introduction, an embedding scheme requires to identify a fundamental unit, the fragment, that is assumed to include all the non-trivial physics arising from the interactions. In our case, where we deal with a full solid with translational invariance, the key concept is to assume that the lattice self-energy is composed by periodic repetitions of a fragment self-energy. 

For the sake of definiteness and simplicity, we start from a single-site fragment and local on-site interactions. Hence, the lattice can be seen as a periodic arrangement of the single-site fragments which are by construction connected only by a single-particle hopping. 
Here, the simplest quantum embedding hypothesis can be summarized in assuming the self-energy to be block-diagonal on the (periodically-repeated) fragment basis and does not depend on their index, i.e. $\braket{\alpha | \Sigma(\omega) | \beta} = \Sigma_{11}(\omega) \delta_{\alpha\beta}$ with $\alpha,\beta$ labeling the periodic replica of each fragment. Eq.~\eqref{eqn:dyson} on this fragment basis becomes:

\begin{widetext}
    \begin{equation}
        \begin{pmatrix}
            \omega + \mu - h_{11} - \Sigma_{11}(\omega) & -h_{12} & \dots \\
            -h_{21} & \omega  + \mu - h_{22} - \Sigma_{11}(\omega) & \dots \\
            \vdots & \vdots & \ddots \\
        \end{pmatrix}
        \begin{pmatrix}
            G_{11}(\omega) & G_{12}(\omega) & \dots \\
            G_{21}(\omega) & G_{22}(\omega) & \dots \\
            \vdots & \vdots & \ddots \\
        \end{pmatrix}
        = \mathbb{I} \, .
        \label{eqn:dyson_Sigma_diag}
    \end{equation}
\end{widetext}




Given the form of Eq.~\eqref{eqn:dyson_Sigma_diag}, 
\begin{equation}
    \begin{pmatrix}
            D_1(\omega) & V \\ V^\dagger & R(\omega)
    \end{pmatrix} 
    \begin{pmatrix}
        G_{11}(\omega) & G_{1R}(\omega) \\ G_{R1}(\omega) & G_{R}(\omega) \\
    \end{pmatrix} = 
    \begin{pmatrix}
            I_{1} & 0 \\ 0 & I_R
    \end{pmatrix}\, ,
    \label{eqn:emb_problem}
\end{equation}
with $D_1=\omega + \mu - h_{11} - \Sigma_{11}(\omega)$, $I_{1}$ the identity on the first site, and the rest evident from the notation, this is a $4\times4$ system of non-commuting matrices. 
Assuming the invertibility of $R(\omega)$~\footnote{Note that this is a very weak assumption as $R(\omega)$ is the Dyson equation for the lattice with one site removed, since we assume that the solution of this system can always be found regardless the number of sites, $R(\omega)$ is invertible.}, the first equation for the system can be re-written as $G_{11}(\omega)^{-1} = D_1(\omega) - V R^{-1}(\omega) V^\dagger$.
By calling  $V R^{-1}(\omega)V^\dagger=v_\text{emb}(\omega)$ the embedding potential, and noting its functional dependence from $R(\omega)$ and $V$, finally we get:

\begin{equation}
    \begin{split}
         G_{11}^{-1}(\omega) & = \omega + \mu - h_{11} - \Sigma_{11}(\omega) - v_\text{emb}[h_{\alpha\beta},\Sigma_{11}](\omega) \, .
    \end{split}
    \label{eqn:emb_selfcons}
\end{equation}

The latter is an embedded Dyson equation arising from the embedding of a fragment with all its periodic replicas.
Crucially, Eq.~\eqref{eqn:emb_selfcons} together with the Dyson equation for the lattice, Eq.~\eqref{eqn:dyson}, give a self-consistent condition for the self-energy $\Sigma_{11}(\omega)$.
Owing to the quantum-embedding hypothesis, the two self-energies $\Sigma_{11}$ appearing in Eq.~\eqref{eqn:emb_selfcons}---the explicit one arising from correlations on the local site $1$, and the other entering through the functional dependence of $v_\text{emb}$ from the contribution of all other sites---must coincide.
Thus, together with a method to solve the embedded system described by Eq.~\eqref{eqn:emb_selfcons}, a given embedding potential $v_\text{emb}$ must generate the same self-energy that was used to construct it---namely, $V R^{-1}(\omega)V^\dagger = v_\text{emb}(\omega)$.
The algorithm to enforce this self-consistency condition and its linking it to existing literature is detailed in Sec.~\ref{sec:selfcons_algo}.

We notice in passing that this formulation, termed here self-consistent dynamical quantum embedding on a lattice, can be easily generalized to any off-diagonal self-energy form---i.e., including inter-fragment terms, such as, $\Sigma_{12}$---~\footnote{Such interfragment self-energy should not be confused with the intersite self-energy within a given fragment.}.
The set of coupled equations within this generalization is beyond the scope of the present work and will be the subject of future investigations.
Below and in the rest of the paper, we focus on the application of the theory within the block diagonal approximation.

It is important to stress that, within the block-diagonal approximation, the self-consistent condition from the coupled Eqs.~\eqref{eqn:emb_selfcons} and~\eqref{eqn:dyson} is equivalent to the condition on the local self-energy of dynamical mean-field theory when the fragment is composed by a single site while the so-called cellular dynamical mean-field theory is obtained from multi-sites fragments. 
Furthermore, the formulation applies directly to solids with strong correlations localized around defect sites. In that case, the derivation yields the same embedded Dyson equation, but without the functional dependence of the embedding potential on the fragment self-energy. Since here we focus on periodic systems, we leave the application of the framework to defects in solids for future work.



\subsection{Solution of the embedded Dyson equation~\eqref{eqn:emb_selfcons} using sum over poles}
\label{sec:emb_dyson}

\begin{figure}[t!]
    \centering
    \includegraphics[width=0.5\textwidth]{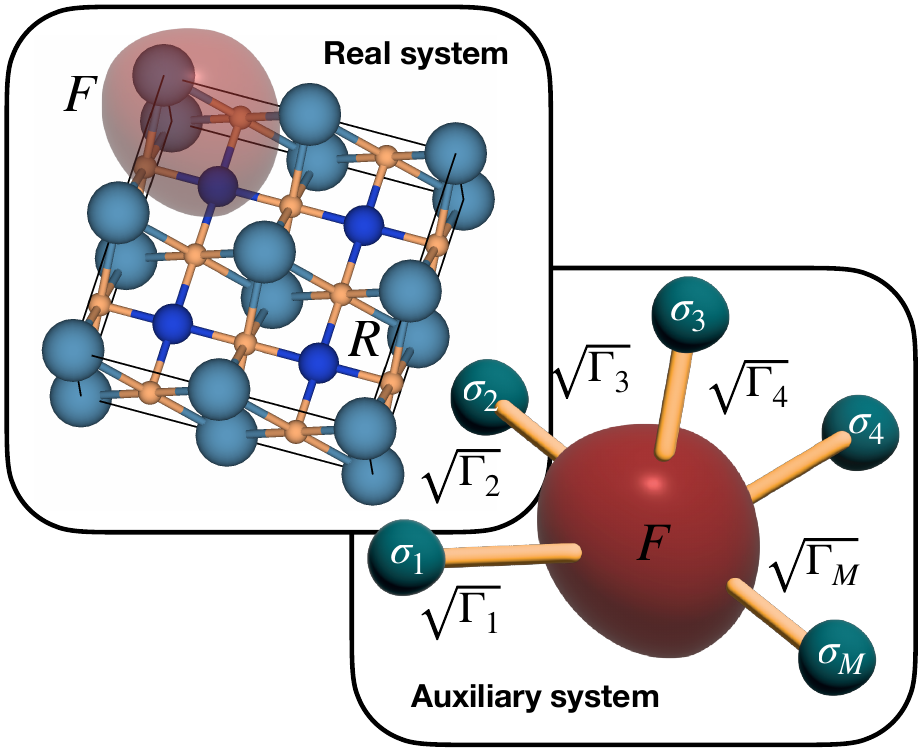}
    \caption{Visual representation of the generalization of the algorithmic-inversion method on sum over poles (SOP), starting from the embedding potential in Eq.~\eqref{eqn:vemb_SOP}. In the top left box a crystal lattice is taken as example of a real periodic system partitioned into the fragment $F$ and its periodic replicas $R$, as in Eq.~\eqref{eqn:emb_problem}. In the bottom right box a sketch of the auxiliary system described by the auxiliary free-energy operator in Eq.~\eqref{eqn:mbAIMSOP_Ham}. The fragment $F$ is surrounded by a collection of non-interacting bath sites exchanging particles only with $F$. The bath sites are parameterized by the poles of the SOP representation in Eq.~\eqref{eqn:vemb_SOP}; the coupling with the fragment by the square root of the residues of the SOP.}
    \label{fig:mbAIMSOP_thm}
\end{figure}

To solve Eq.~\eqref{eqn:emb_selfcons}, we generalize the algorithmic-inversion method on sum over poles to the many-body case.
Specifically, by a sum-over-poles (SOP) representation of the embedding potential we build an auxiliary many-body Hamiltonian that is constrained to have the same local Green's function (GF) of the embedded Dyson equation above.
We start by defining:

\begin{equation}
    v_\text{emb}(\omega) = \sum_{m=1}^M \frac{\Gamma_m}{\omega - \sigma_m} \, , 
    \label{eqn:vemb_SOP}
\end{equation}

where $\Gamma_m$ is a complex matrix and $\sigma_m$ is a complex scalar. 
Please note that, by construction, $\Gamma_m$ acts on the fragment, leaving the internal degrees of freedom of the matrix determined by its rank: the matrix factorization, e.g., using a singular value decomposition, $\Gamma_m=V_m\bar{V}^\dagger_m$, yields rows (columns) of $V_m$ ($\bar{V}^\dagger_m$) spanning the fragment, while leaving the columns (rows) of $V_m$ ($\bar{V}^\dagger_m$) as internal DOFs of the mapping.
The auxiliary free-energy operator that satisfies the constraint on the local GF is:

\begin{align}
    \begin{split}
        \hat{K}^\text{aux} & = \hat{K}_{F} + 
        \sum_{m=1}^M \left(\hat{c}^\dagger_{\zeta} [V_m]^{\zeta\nu_m}  \hat{d}_{\nu_m} 
        + \hat{d}^\dagger_{\nu_m} [\bar{V}^\dagger_m]^{\nu_m\zeta} \hat{c}_{\zeta} \right) \\ 
        & + \sum_{m=1}^M \sigma_m \hat{d}^\dagger_{\nu_m} \hat{d}_{\nu_m} \, .
    \end{split}
    \label{eqn:mbAIMSOP_Ham}
\end{align}

where $\hat{K}_{F} = \hat{H}_F - \mu \hat{N}_F$ is the free-energy operator of the fragment, $\zeta$ labels the DOF of the fragment ($F$), $\nu_m$ spans the auxiliary internal DOF for each pole $m$, and repeated indexes labeling internal DOF are summed implicitly.
Here, $\hat{c}$ is the fermionic operator of the fragment and $\hat{d}$ is a fermionic operator in the auxiliary system that we refer to as (fictitious) bath.
Note that for simplicity of notation we have not labeled $\hat{d}$ as $\hat{d}^{(m)}$ for each pole $m$, as the $m$ index for $\hat{d}$ is already evident from its internal DOF $\nu_m$.
As said above, these auxiliary DOF spanned by $\nu_m$ are determined by the rank of $\Gamma_m$.

Eventually, it is possible to show that the local Green's function $G_{F}^\text{aux}=P_FG^\text{aux}P_F$ generated by $\hat{K}^\text{aux}$ satisfies the embedded Dyson equation~\eqref{eqn:emb_selfcons} for a given embedding potential.
We prove this equivalence in Appendix~\ref{app:mbAIMSOP}.
In Fig.~\ref{fig:mbAIMSOP_thm} we depict pictorially this mapping between the embedding of the Dyson equation on a single fragment from its periodic replicas and the auxiliary system to solve the equation.
In particular, thanks to the formal expression of the auxiliary Hamiltonian, the method enables the evaluation of the local Green’s function for a fixed embedding potential without solving the Dyson equation in Eq.~\eqref{eqn:emb_selfcons} at each frequency.

Here, it is important to stress that the aforementioned mapping to an auxiliary system is equivalent to an impurity model with a discretized bath within a DMFT framework.
In this work we limit the treatment to the exact diagonalization (ED) of the auxiliary Hamiltonian (or free energy) in Eq.~\eqref{eqn:mbAIMSOP_Ham}, similarly to ED solvers in DMFT, but our approach can be combined also with tensor-network-based methods, e.g. matrix product states (MPS) and density matrix renormalization group (DMRG) \cite{Capone,senechal_2000, wolf_2014, nishimoto_2006,bauernfeind_2017}.  



\subsection{Self-consistent algorithm}
\label{sec:selfcons_algo}

In this Section we explicit the implementation that we use to achieve the self-consistent condition outlined in Sec.~\ref{sec:quantum_emb_lat}.
Letting ${(i)}$ be the iteration index, given a chemical potential $\mu^{(i)}$ and a SOP representation of the embedding potential $v^{(i)}_\text{emb}$---either given or coming from a previously calculated self-energy $v^{(i)}_\text{emb}[\Sigma^{(i-1)}_{11}]$---we start by solving Eq.~\eqref{eqn:mbAIMSOP_Ham} to find the local Green's function of the fragment ${G}_F^{(i)}(\omega)$.
As this step is identical to the impurity part of a DMFT procedure, we take advantage of the machinery used for solvers of DMFT and employ an exact diagonalization method.
Specifically, here we employ a Lanczos methodology that is outlined in Appendix~\ref{app:bilanczos}.
As said in Sec.~\ref{sec:emb_dyson} and showed in Appendix~\ref{app:mbAIMSOP}, ${G}_F^{(i)}$ satisfies the embedded Dyson equation~\footnote{Note that as $v^{(i)}_\text{emb}[\Sigma^{(i-1)}_{11}]$, Eq.~\eqref{eqn:selfconst_cycle} is different than the self-consistent equation~\eqref{eqn:emb_selfcons}, exactly bringing the difference between $G_{F}$ and the local Green's function $G_{11}$ if not at self consistency.}:
\begin{equation}
    \begin{split}
         G_{F}^{(i)}(\omega) & = \left[\omega + \mu^{(i)} - h_{11} - \Sigma^{(i)}_{11}(\omega) - v^{(i)}_\text{emb}(\omega)\right]^{-1} \, .
    \end{split}
    \label{eqn:selfconst_cycle}
\end{equation}
Thus, we can use the equation above to find the local self-energy $\Sigma^{(i)}_{11}$.
We can now solve the Dyson equation~\eqref{eqn:dyson} to get the Green's function of the system $G^{(i)}$, and localize it to arrive to $G^{(i)}_{11}$.
As we are on a lattice, these three steps---namely unfolding of the self-energy, the solution of the Dyson equation, and localization---can be done simply by:
\begin{equation}
    G^{(i)}_{11}(\omega) = \frac{1}{N_\mathbf{k}} \sum_\mathbf{k} \bigl( \omega + \mu^{(i)} - \varepsilon_\mathbf{k} - \Sigma^{(i)}_{11}(\omega) \bigr)^{-1} \, .
    \label{eqn:Gloc_DMFT}
\end{equation}
To find the updated embedding potential we can use its definition from Sec.~\ref{sec:quantum_emb_lat}, $G^{(i)}_{11}(\omega)^{-1} = D^{(i)}_1(\omega) - v^{(i)}_\text{emb}(\omega)$, which explicitly reads:
\begin{equation}
    v^{(i+1)}_\text{emb}(\omega) = \omega + \mu^{(i)} - h_{11} -  G^{(i)}_{11}(\omega)^{-1} - \Sigma^{(i)}_{11}(\omega) \, .
    \label{eqn:vemb_DMFT}
\end{equation}
The updated embedding potential $v^{(i+1)}_\text{emb}[\Sigma^{(i)}_{11}]$ can be used in a new cycle.
Furthermore, to update the chemical potential we enforce the conservation of the number of particles by asking $N=\int^{\mu^{(i+1)}}_{-\infty}d\omega\, e^{i\omega0^+}\mathrm{Tr}[G^{(i)}(\omega)]$.
Finally, again in perfect parallel to DMFT, to find the SOP representation of $v^{(i+1)}_\text{emb}(\omega)$ we minimize the cost function on a frequency representation:
\begin{equation}
    C\left( \{ (\Gamma_m, \sigma_m) \}_{m=1}^M \right) = \int d\omega  \left\lVert v^{(i+1)}_\text{emb}(\omega) - \sum_{m=1}^M \frac{\Gamma_m}{\omega - \sigma_m} \right\rVert^2 \, ,
    \label{eqn:cost_func_DMFT}
\end{equation}
which gives a new set of residues $\Gamma_m^{(i+1)}$ and poles $\sigma_m^{(i+1)}$ of Eq.~\eqref{eqn:vemb_SOP} that, together with $\mu^{(i+1)}$, allow us to define $\hat{K}^{\text{aux}(i+1)}$ in Eq.~\eqref{eqn:mbAIMSOP_Ham} and repeat the cycle.

Performing this minimization and in general the whole self-consistent procedure directly on real frequencies would give direct access to spectral resolution, but it is impeded by the non analytical character of the time-ordered propagators.
Therefore, most DMFT implementations are formulated on the imaginary-frequency axis, meaning that all dynamical quantities are computed on an imaginary grid.
While this is ideal to compute thermodynamic (frequency-integrated) quantities and address their temperature dependence, the prediction of spectral resolved quantities, such as the density of states or angle-resolved photoemission, requires analytic continuation procedures.
Although these have been developed extensively for this task~\cite{fei_2021,han_analytic_2017}, here we propose to alleviate this problematic by simply performing the self-consistent cycle directly ``near'' the real axis.
Specifically, we introduce the so-called erf path:
\begin{equation}
    z(\omega) = \omega + i \eta \text{erf}(\omega) \, ,
    \label{eqn:erf_axis}
\end{equation}
where $\eta$ is a positive parameter modulating the distance from the real frequencies. 
Notably, within this choice $\text{Im}z(\omega) \to \eta$ when $\omega \to \infty$ and $z(\omega)\to\omega$ when $\eta\to0^+$. In particular, all our quantities are evaluated on a discrete mesh with a controlled number of points $N$.

Aside the frequency representation below, we also introduce a different strategy to perform the self-consistent cycle addressing the minimization of the cost function in Eq.~\eqref{eqn:cost_func_DMFT}.
As evidenced for the procedure above, the whole self-consistent procedure can be viewed as a update for the poles and residues of the embedding potential so called SOP.
This suggests an alternative scheme to achieve self consistency based on a smooth update of this SOP representation, that we refer to as dynamics of poles and residues.
Consistently with Eq.~\eqref{eqn:cost_func_DMFT}, we move poles and residues using the gradient of the cost function namely:
\begin{equation}
    \begin{cases}
    \Delta\Gamma_m=-\lambda\nabla_{\Gamma_m}C \\
    \Delta\sigma_m=-\lambda\nabla_{\sigma_m}C
    \end{cases},
    \label{eqn:SD_respoles}
\end{equation}
with $\lambda$ a chosen positive parameter which is constant throwout the self-consistent cycle.
Practically, instead of performing a full minimization of the cost function in Eq.~\eqref{eqn:cost_func_DMFT}, we evolve the residues and poles with the single-step case specified in Eq.~\eqref{eqn:SD_respoles}.
When performing multiple steps for the minimization, we implement both a steepest descent (SD) algorithm and a conjugate gradient (CG) procedure from the Polak-Ribière method~\cite{nocedal_2006}.

\subsection{Implementation on the Hubbard ring}

We apply the self-consistent dynamical quantum embedding on a lattice to the case of the one-dimensional Hubbard model with periodic boundary conditions (ring) at zero temperature at half filling. 
The Hubbard Hamiltonian reads:
\begin{equation}
    \begin{split}
        \hat{H} = & t \sum_{i \sigma} \left(\hat{c}_{i\sigma}^\dagger \hat{c}_{i+1\sigma} + \hat{c}_{i+1\sigma}^\dagger \hat{c}_{i\sigma}\right)  +
        U \sum_{i} \hat{n}_{i\uparrow} \hat{n}_{i\downarrow}
    \end{split},
 \end{equation}
with the usual convention that $\sigma$ labels the spin DOF and $\hat{n}_{i\uparrow}$ is number of particles on the site $i$ with spin $\sigma$.
In practice, we study a $1000$ site ring to approximate the thermodynamic limit at half-filling.
Note that the half-filling condition with $U\neq0$ enforces the chemical potential to be $\frac{U}{2}$ analytically, and consequently can be taken fixed during the self-consistent cycle.
Furthermore, we model the paramagnetic (non-magnetic) solution of the model by enforcing the Green's functions (and all the dynamical quantities) to be diagonal and equal in the spin index, $G_\uparrow(\omega) = G_\downarrow(\omega)$. 

With these conditions, the internal dimension of the fragment $\zeta$ in Eq.~\eqref{eqn:mbAIMSOP_Ham} is the spin index $\sigma$, and the embedding potential is proportional to the identity, with its residues reading $[\Gamma_m]_{\sigma\sigma'}=\gamma_m\delta_{\sigma\sigma'}$.
Consequently, $V_m$ is also proportional to the identity as $[V_m]^{\sigma\sigma'}=\sqrt{\gamma_m}\delta_{\sigma\sigma'}$, and the same for $\bar{V}^\dagger_m$.
Indeed, in this simple case were matrices are of maximum rank (identity), $\nu_m=\sigma$, i.e., the auxiliary internal DOF of each pole is equivalent to a spin.
Furthermore, due to the spin restriction, the minimization of the cost function in Eq.~\eqref{eqn:cost_func_DMFT} is equivalent to minimizing the mean-square difference of its scalar arguments, i.e., $v_\text{emb}^{\sigma\sigma}$ and $\Gamma^m_{\sigma\sigma}$, with the result valid for both spin components.
Note that, since we also use real poles together with positive scalar residues, the Hamiltonian of the fragment in Eq.~\eqref{eqn:mbAIMSOP_Ham} is Hermitian.

We perform the study using different values of the interaction, namely $U/t=1,4,7$ and for different numbers of residues/poles.
Regarding the proximity to the real axis, we carry out simulations for $\eta/t=2,1.5,1,0.5$ and compare them to the calculations on the imaginary axis.
The number of frequencies used in the calculation is $10^5$, equally spaced in the range $[-10,10]$.
As said previously, the fragment Hamiltonian is solved by exact diagonalization in close analogy to for example in~\cite{Caffarel_exact_1994,amaricci_2022,Capone,senechal_2000,Crippa2025}.
In particular, here we implement a so-called bi-Lanczos method~\cite{bhattacharya_krylov_2023} that is described in Appendix~\ref{app:bilanczos}.
Please also note that we also check the robustness of the results of self-consistent cycle with respect to the choice of the initial position of poles and residues.

This implementation is distributed as a self-contained open-source Python package available at the GitHub page of the corresponding author~\cite{carbone_code}. The package provides routines to construct sum-over-poles (SOP) representations of Green’s functions and related dynamical functions, perform the associated fitting via conjugate-gradient minimization, and run the dynamical embedding self-consistency loop with the aforementioned exact-diagonalization-based impurity solvers. The code is designed to operate directly on compact rational representations, avoiding explicit analytic continuation.


\section{Results}

\subsection{Benchmark and results on the real axis}
\label{sec:benchmark_real_axis}

\begin{figure}
    \centering
    \includegraphics[width=0.5\textwidth]{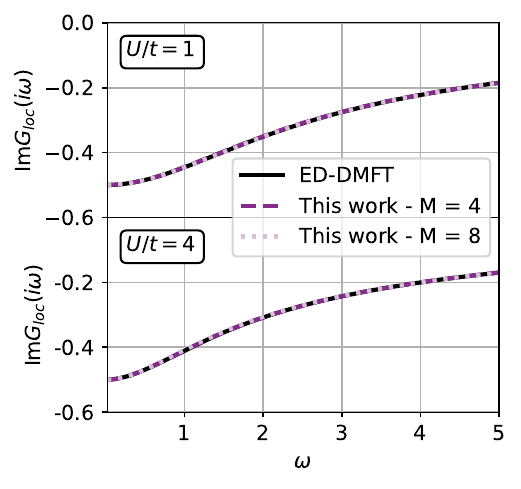}
    \caption{Imaginary parts of the diagonal elements of the local Green's function $G_\text{loc}$ of the half-filled $1000$-site Hubbard ring with $U/t=1$ (up) and $U/t=4$ (down). The reference results from ED-DMFT \cite{capone_2004} are obtained using $M=8$ bath sites.}
    \label{fig:Gloc_imag_comparison}
\end{figure}

\begin{figure*}
    \centering
    \includegraphics[width=\textwidth]{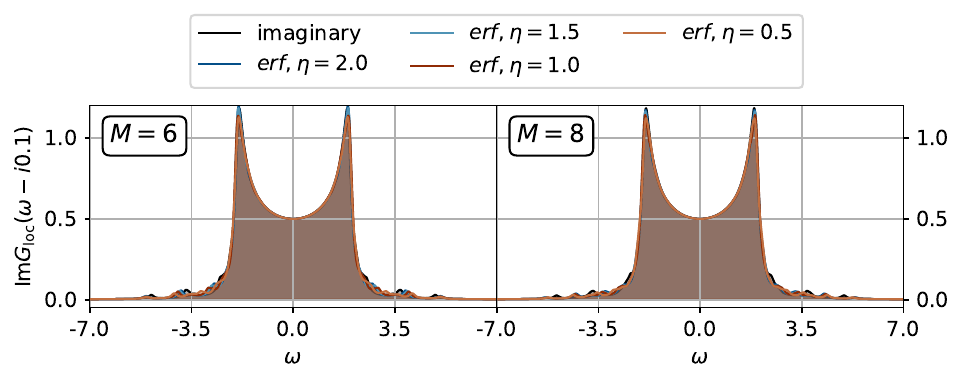}%
    \begin{picture}(0,0)
      \put(-285,0){\large\bfseries (a) $U/t=1$}
    \end{picture}
    \includegraphics[width=\textwidth]{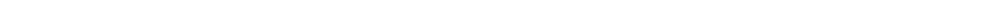} 
    \includegraphics[width=\textwidth]{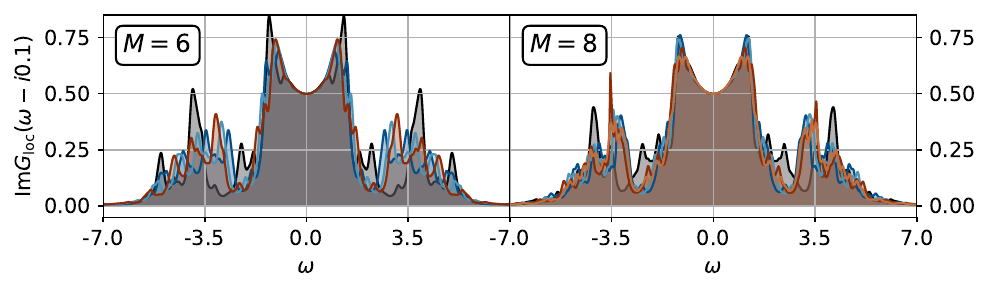}%
    \begin{picture}(0,0)
      \put(-285,0){\large\bfseries (b) $U/t=4$}
    \end{picture}
    \includegraphics[width=\textwidth]{Pictures/white_space.pdf} 
    \includegraphics[width=\textwidth]{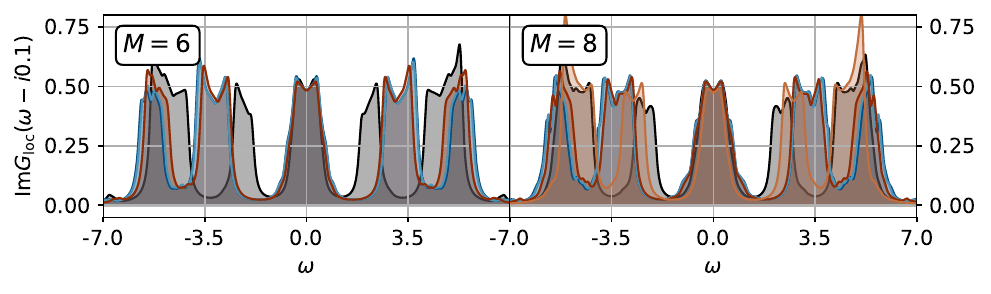}%
    \begin{picture}(0,0)
      \put(-285,0){\large\bfseries (c) $U/t=7$}
    \end{picture}
    \caption{We study the half-filled $1000$-site Hubbard ring at $U/t=1,4,7$ in \textbf{(a)}, \textbf{(b)} and \textbf{(c)}, respectively, with $M=6,8$ residues/poles. We compare the imaginary parts of the local Green's function on the axes $\omega - i \lambda$ with $\lambda = 0.1$. Every curve refers to the DMFT simulation performed either on the imaginary axis (black curve) or on $\omega + i \eta \text{erf}(\omega)$ for different values of $\eta$ (colored curves).}
    \label{fig:Gloc}
\end{figure*}

\begin{figure*}
    \centering
    \subfloat[]{\includegraphics[width=0.49\textwidth]{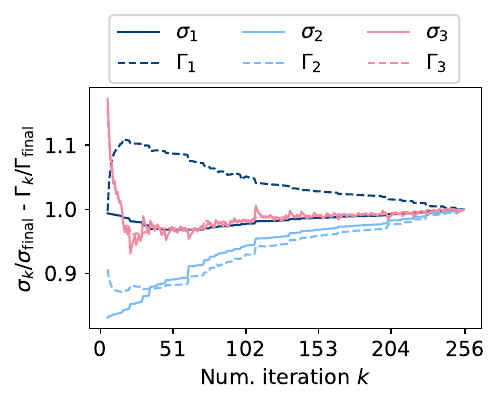}}%
    \begin{picture}(0,0)
      \put(-170,-15){\large\bfseries (a) Dynamics - $M=6$}
    \end{picture}
    \subfloat[]{\includegraphics[width=0.49\textwidth]{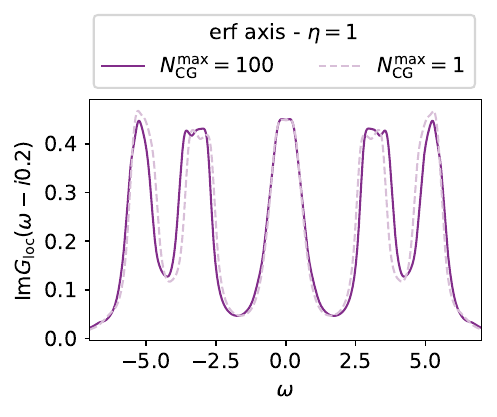}}
    \begin{picture}(0,0)
      \put(-190,-15){\large\bfseries (b) $G_\text{loc}$ comparison - $M=6$}
    \end{picture} 
    \\
    \subfloat[]{\includegraphics[width=0.49\textwidth]{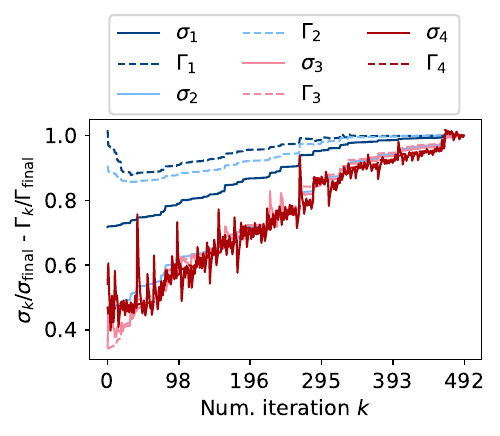}}%
    \begin{picture}(0,0)
      \put(-170,-15){\large\bfseries (c) Dynamics - $M=8$}
    \end{picture}
    \subfloat[]{\includegraphics[width=0.49\textwidth]{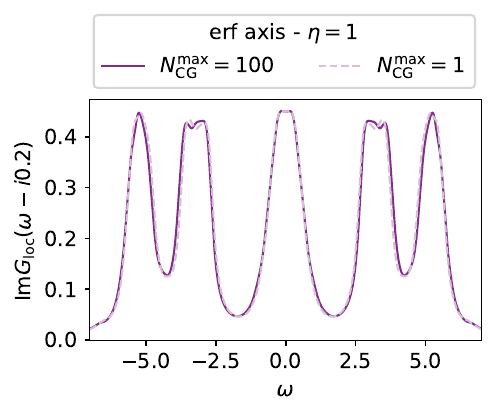}}
    \begin{picture}(0,0)
      \put(-190,-15){\large\bfseries (d) $G_\text{loc}$ comparison - $M=8$}
    \end{picture} 
    \caption{Comparison of the "pole and residue" dynamics methodology against the reference results of Sec.~\ref{sec:benchmark_real_axis} for the half-filled Hubbard ring simulation at $M=6,8$ and $U/t = 7$ on the erf axis with $\eta=1$. In \textbf{(a)} and \textbf{(c)} we show the evolution (trajectory) of the parameters of the SOP representation using the ratio of each residue/pole with the reference---$\sigma_k/\sigma_\text{Ref}$, and $\Gamma_k / \Gamma_\text{Ref}$---at $M=6,8$, respectively. As at half filling the poles and residues are symmetric we show the ones pertaining to the positive axis. In \textbf{(b)} and \textbf{(d)} the converged local Green's functions $G_\text{loc}$ are analytically continued to the complex axis $\omega - i 0.2$ and compared with the reference.}
    \label{fig:dyn_poles}
\end{figure*}

To benchmark the implementation, we note that, when applied to the Hubbard Hamiltonian and performed on the imaginary axis, the self-consistent quantum embedding procedure outlined in this work is identical to the exact diagonalization implementation of DMFT.
In Fig.~\ref{fig:Gloc_imag_comparison} we compare the results taken from the implementation of this work to those from Ref.~\cite{capone_2004}, for different values of the interaction and dimension of the fragment or impurity.
The results are in perfect agreement to ED-DMFT on the imaginary axis, giving proof of the correctness of the implementation.



\bigskip

We now move to the simulations close to the real axis. 
Specifically, we present the results of our simulations on the erf axis in Eq.~\eqref{eqn:erf_axis} with $\eta/t$ ranging from $0.5,1,1.5,2$ using $M=6,8$ poles and with $U/t=1,4,7$. 
In Fig.~\ref{fig:Gloc} we show the imaginary parts of the local Green's functions $G_\text{loc}(\omega - i\lambda)$, where the broadening parameter is set to $\lambda/t=0.1$ in order to highlight spectral details. 
Panel (a) of Fig.~\ref{fig:Gloc} shows the simulation at $U/t=1$ for $M=6$ (left) and $M=8$ (right) number of poles, comparing the result at the different $\eta$ values.
As expected, since the interaction is small, there is little dependence from any of these parameters showing two dominant peaks at $\omega/t=\pm 2$ in all simulations. 
Panel (b) of Fig.~\ref{fig:Gloc} reports the simulation at $U/t=4$ with left and right plots with $6$ and $8$ poles.
For this value of the interaction, a dependence on $\eta$ becomes evident for $M=6$ and reduced but still visible for $M=8$.
In particular the low-energy features--- $\omega/t$ between $0$ and $2$---are weakly dependent on both $\eta$ and $M$ and the differences are confined in the high-energy part of the spectrum.
This trend is confirmed in panel (c) of Fig.~\ref{fig:Gloc} where differences in the high-energy features are even more pronounced for both $\eta$ and $M$.

\subsection{Pole and residue dynamics}

In this Section we implement numerically the "pole and residue" dynamics introduced in Eq.~\eqref{eqn:SD_respoles}.
Specifically, we want to compare the result of the previous Section on the real axis, without minimizing the cost function $C$ in Eq.~\eqref{eqn:cost_func_DMFT} but following only the force, i.e., minus the gradient of $C$, at each self-consistent cycle.
Here, we focus on the correlated case $U/t=7$ but report that for the other cases very similar conclusions were drawn.
Also, we carry out this investigation for two different numbers of poles, $M=6$ and $M=8$, at $\eta/t=1$.
In panels (b) and (d) of Fig.~\ref{fig:dyn_poles} we compare the two cases against the reference found in the previous Section.
Even though the $M=6$ simulation is qualitatively comparable, only with $M=8$ we get quantitative agreement.
The reason is the more pronounced $\eta$ dependence at $M=6$ than at $M=8$, which testifies to the reduced stability of the $M=6$ simulation at $\eta/t=1$.
Panels (a) and (c) show the evolution of poles and residues during the self-consistent cycle with the "pole and residue" dynamics method against the reference result obtained in the section before.
Focusing on the poles for simplicity, with $M=6$ the poles only land qualitatively on the reference ones, while with $M=8$ the result is again quantitative.
Other than confirming the results of the previous Section, the results of the "pole and residue" dynamics investigation are two-fold: first, they show that the dynamics of poles and residues can be run smoothly only using the force on them, i.e., minus the gradient of the cost function in Eq.~\eqref{eqn:cost_func_DMFT}; second, that only at convergence we can expect that the result from the standard algorithm which includes the minimization of $C$, and the present one, coincide.
Although beyond the scope of the present work, the former conclusion can be used in noisy simulations where the convergence is difficult to reach, e.g., the case of alternation of quasi-degenerate magnetic states encountered in Refs.~\cite{Peters_2009,Yu_2023}; the latter can be used (as we did here) to check the robustness with respect to the optimization path, as the full-minimization reference method and the present one have very different dynamics but coincide at convergence. Overall, the "pole and residue" dynamics emerges as a viable alternative to the full minimization of $C$ at each self-consistent cycle, provided that the number of poles $M$ is large enough to guarantee stability at the chosen broadening $\eta$.

\section{Conclusion}

We have introduced a dynamical quantum embedding framework for many-body lattice Hamiltonians on a sum-over-poles (SOP) representation of the embedding potential.
Starting from the lattice Dyson equation, we build an embedding scheme by identifying a periodically-repeated fragment that in the simplest case treated here is the site. 
With this approximation we derive an embedded Dyson equation for the fragment, Eq.~\eqref{eqn:emb_selfcons}, that together with the Dyson equation for the lattice, Eq.~\eqref{eqn:dyson}, form a set of self-consistent equations coupled by the embedding potential $v_\text{emb}(\omega)$.
By writing the embedding potential $v_\text{emb}(\omega)$ as a SOP, we map it onto an auxiliary many-body system---the single interacting fragment coupled to a finite non-interacting bath. 
This generalizes the algorithmic-inversion method to the interacting case for a real-space embedding and reproduces the discretized-bath impurity model underlying exact-diagonalization implementations of dynamical mean-field theory (DMFT), with the bath levels and couplings set, respectively, by the poles and the square-root residues of the SOP.

Motivated by the embedding construction, we implement the self-consistent cycle---impurity solution, lattice upfolding, and fit of the embedding potential---close to the real-frequency axis, along the erf path $z(\omega)=\omega+i\eta\,\text{erf}(\omega)$. Since the auxiliary Hamiltonian delivers the Green's function on arbitrary complex-frequency contours without solving the Dyson equation frequency by frequency, real-frequency spectral quantities are accessed directly, avoiding ill-posed analytic continuation procedures from the imaginary axis.

We benchmarked the method on the half-filled one-dimensional Hubbard ring at zero temperature for $U/t=1,4,7$ and bath sizes $M=6,8$. On the imaginary axis the results coincide with ED-DMFT, validating the implementation; on the erf axis, these compact expansions already reproduce the dominant spectral features. The dependence on the contour distance $\eta$ and on the number of poles $M$ is concentrated in the high-energy part of the spectrum and grows with the interaction strength, while the low-energy features remain robust.

We further recast the self-consistency as a dynamics of the SOP parameters, evolving poles and residues along the force $-\nabla C$ of the fitting cost function. A single such step per self-consistency cycle already reaches the same converged solution obtained by fully minimizing the cost function, showing that a complete fit is not a prerequisite for self-consistency: the embedding loop itself drives the parameters toward the solution. This well-behaved update of the SOP representation can be useful when convergence is hard to reach or demonstrate independence from the optimization path.

Overall, these results establish the SOP representation as a practical route to real-frequency quantum embedding. Beyond the single-site, single-orbital case studied here, the construction extends naturally to multi-orbital and realistic electronic-structure Hamiltonians, to cluster fragments with inter-fragment self-energies, and---through the variational view of poles and residues---to self-consistency algorithms beyond standard DMFT.

\appendix

\section{Embedded Dyson equation and auxiliary local quantities}
\label{app:mbAIMSOP}

In this Section we show that the lattice and the auxiliary systems share the same local Green's function provided that they also share the same embedding and chemical potentials $v_\text{emb}(\omega)$ and $\mu$.
We start by considering the embedded Dyson equation of the lattice, Eq.~\eqref{eqn:emb_selfcons}, and note that in a functional theory---i.e. where the self-energy is the derivative of a (Klein or Luttinger-Ward) functional---that is also diagrammatic, the simple quantum embedding hypothesis we made at the beginning of the derivation, i.e. $\Sigma_{\alpha\beta}(\omega)=\Sigma_{11}(\omega)\delta_{\alpha\beta}$, can be translated in the skeletonic diagrams of local self-energy $\Sigma_{11}$ having only lines inside the fragment (or any of its periodic replicas)~\footnote{This follows directly from the functional definition of the self-energy. Under the current hypotheses we have $\frac{\delta\Phi}{\delta G_{\alpha\beta}}=0$ if $(\alpha,\beta)$ $\notin$ the fragment, and thus $\Phi$ independent (or constant) from $G$ outside the fragment, or equivalently $\Phi[G_{11}]$.}.
Equivalently, this means $\Sigma_{11}[G_{11}]$.
Thus, we can rewrite the embedded Dyson Equation~\eqref{eqn:emb_selfcons} for the lattice as:
\begin{equation}
    \begin{split}
         G_{11}^{-1}(\omega) & = \omega + \mu - h_{11} - \Sigma_{11}[G_{11}](\omega) - v_\text{emb}(\omega) \, .
    \end{split}
    \label{eqn:emb_selfcons_Gdepend}
\end{equation}
In parallel we note that the local Green's function of the auxiliary system satifies a formally-equivalent embedded Dyson equation.
This has been already shown in Appendix~A3 of Ref.~\cite{ferretti_2024} and derives directly from $\hat{K}^\text{aux}$ in Eq.~\eqref{eqn:mbAIMSOP_Ham}, where the interaction two body term is restricted to be only in $H_F$. 
Indeed, as  the two body term in $H_F$ has only operators in the fragment (here there are no periodic replicas as we are in the auxiliary fragment+bath system) the self-energy is nonzero only in the fragment, i.e., $\Sigma^\text{aux}=\Sigma_F^\text{aux}$ inside the fragment and zero otherwise.
Then, by rewriting Eq.~\eqref{eqn:emb_problem} with the self-energy present only in the fragment (first element for simplicity), the auxiliary system satisfies the same Dyson equation~\eqref{eqn:emb_problem} now with $D_1(\omega)=\omega+\mu-h_{11}-\Sigma_F^\text{aux}(\omega)$ ($h_F=h_{11}$ by construction), the coupling $V=\begin{pmatrix} V_1, & \cdots, & V_M \end{pmatrix}$ (with $V^\dagger$ the corresponding column of blocks $\bar{V}_m^\dagger$), and the bath block $R(\omega)=\bigoplus_{m=1}^M(\omega-\sigma_m)\,I_{\nu_m}$ (here the direct sum stacks the blocks block-diagonally), independent from the self-energy.
Again, by the invertibility of $R(\omega)$ the projection on the fragment of the Green's function $G_F^\text{aux}=P_F G_F^\text{aux} P_F$ satisfies:
\begin{equation}
    \begin{split}
         G_{F}^{-1}(\omega) & = \omega + \mu - h_{11} - \Sigma_F^\text{aux}(\omega) - v_\text{emb}(\omega) \, ,
    \end{split}
    \label{eqn:emb_auxiliary}
\end{equation}
where the embedding potential is the same as the one of the real system, as
\begin{equation}
    v_\text{emb}^\text{aux}(\omega) = V R^{-1}(\omega)\, V^\dagger = \sum_{m=1}^M V_m\,\frac{1}{\omega-\sigma_m}\,\bar{V}_m^\dagger = v_\text{emb}(\omega) \, ,
    \label{eqn:vemb_aux}
\end{equation}
where we used the diagonal form of $R(\omega)$ and the factorization of the residues $\Gamma_m=V_m\bar{V}_m^\dagger$ introduced in Eq.~\eqref{eqn:vemb_SOP}.
In addition, and following~\cite{ferretti_2024}, the strictly local two-body term of $\hat{K}^\text{aux}$ generates self-energy diagrams with vertices only inside the fragment, so that the auxiliary self-energy is a functional of the local Green's function alone, $\Sigma_F^\text{aux}=\Sigma_F^\text{aux}[G_F^\text{aux}]$, exactly as $\Sigma_{11}=\Sigma_{11}[G_{11}]$ for the lattice.
Resolving the skeletonic dependence, i.e.\ re-expanding the dressed local propagators in terms of the non-interacting ones~\cite{ferretti_2024,martin_reining_ceperley}, each self-energy becomes a functional of the corresponding non-interacting local propagator, $\Sigma_{11}=\Sigma_{11}[G^0_{11}]$ and $\Sigma_F^\text{aux}=\Sigma_F^\text{aux}[G^{0,\text{aux}}_F]$; since the local two-body interaction is identical in the two systems, the local diagrams---and hence the functional---coincide.
Now, since Eqs.~\eqref{eqn:emb_selfcons_Gdepend} and~\eqref{eqn:emb_auxiliary} share $h_{11}$ and, by Eq.~\eqref{eqn:vemb_aux}, the same embedding potential, the two non-interacting local propagators coincide, $G^0_{11}=G^{0,\text{aux}}_F=[\,\omega+\mu-h_{11}-v_\text{emb}(\omega)\,]^{-1}$.
The same functional evaluated on the same $G^0$ then yields $\Sigma_{11}(\omega)=\Sigma_F^\text{aux}(\omega)$, and therefore $G_{11}(\omega)=G_F^\text{aux}(\omega)$, which finally proves the equivalence.

\section{Exact diagonalization of the fragment using bi-Lanczos}
\label{app:bilanczos}
For values of $M >6$ we use the Lanczos algorithm \cite{dagotto_1994} when generalized to the so-called bi-Lanczos method \cite{bhattacharya_krylov_2023}. If two distinct right and left vectors, say $\ket{\phi_0}$ and $\bra{\chi_0}$, are given, the following expression for a generic Hamiltonian $\hat{H}$ holds:

\begin{equation}
    \bra{\phi_0} \bigl( \omega - \hat{H} \bigr)^{-1} \ket{\chi_0} = \frac{\braket{\phi_0 | \chi_0}}{\omega - \alpha_1 - \frac{\beta_2 \gamma_2}{\omega - \alpha_2 - \dots}} \, ,
    \label{eqn:bilanczos}
\end{equation}

where $\alpha_k,\beta_k,\gamma_k$ are the non-zero elements of the tridiagonal matrix expressed on the bi-orthogonal bi-Lanczos bases $P = (p_0 p_1 \dots p_m \dots )$ and $Q = (q_0 q_1 \dots q_m \dots)$:

\begin{equation}
    Q^\dagger \hat{H} P =    
    \begin{pmatrix}
        \alpha_0 &\beta_1 & 0 &  &  &0 \\
        \gamma_1 & \alpha_1 & \beta_2 & & & \\
        0 &\gamma_2 & \alpha_2 &\ddots & & \\
        & &\ddots &\ddots &\beta_m & \\
        & & &\gamma_m & \alpha_m &\ddots \\
        0 & & & &\ddots &\ddots \\
    \end{pmatrix} \, .
    \label{eqn:tridiag_mat}
\end{equation}

Once the chemical potential is fixed, the ground state $\ket{E_0}$ of the Hamiltonian in Eq.~\eqref{eqn:mbAIMSOP_Ham} in the full-fermionic Fock space is calculated efficiently using sparse-based diagonalization routines. The left and right vectors derived from the ground state of $\hat{K}^\text{aux}$, $\bra{E_0}\hat{O}$ and $\hat{O}'\ket{E_0}$ with $\hat{O}, \hat{O}' = \hat{c}_i, \hat{c}_i^\dagger$, are now used as $\bra{\phi_0}$ and $\ket{\chi_0}$ in Eq.~\eqref{eqn:bilanczos} to evaluate the electron addition and removal terms of the time-ordered Green's function following the bi-Lanczos algorithm. 

\bigskip

\bibliography{main}

\end{document}